\documentclass[12pt,reqno]{amsart}
\usepackage{eurosym}
\usepackage{hyperref}
\usepackage{appendix}
\usepackage{graphicx}
\usepackage{amscd}
\usepackage{amsmath}
\usepackage{epsfig}
\usepackage{amsfonts}
\usepackage{amssymb}
\usepackage{setspace}
\usepackage{listings}
\usepackage{framed}
\usepackage{xcolor}

\colorlet{shadecolor}{gray!20}
\providecommand{\U}[1]{\protect\rule{.1in}{.1in}}
\providecommand{\U}[1]{\protect\rule{.1in}{.1in}}
\allowdisplaybreaks

\theoremstyle{plain}

\numberwithin{equation}{section}

\input{tcilatex}
\newcommand{\grayitss}[1]{\textcolor{gray}{\textsf{\textit{#1}}}}
\lstdefinestyle{matitalic}{
  language=Mathematica,
  mathescape=true,
  basicstyle=\ttfamily\small,
  columns=fullflexible,
  keepspaces=true,
  breaklines=true,
  upquote=true,
  literate=
    {In[}{{\grayitss{In[}}}1
    {]:=}{{\grayitss{]:=}}}1
    {Out[}{{\grayitss{Out[}}}1
    {]=}{{\grayitss{]=}}}1,
 }
\lstnewenvironment{mat}{\lstset{style=matitalic}}{}

\begin{document}
\title[Classification of Transuranium Elements]{Classification of
Transuranium Elements in Terms of `Winding' Numbers in the Bohr-Sommerfeld
Model }
\author{Sergei K. Suslov}
\address{School of Mathematical and Statistical Sciences, Arizona State
University, Tempe, AZ 85287--1804, U.S.A.}
\email{sks@asu.edu}
\date{November 17, 2026}

\begin{abstract}
We utilise the Bohr-Sommerfeld atomic model to explore
hydrogen-like ions from Uranium ($Z=92$) to Oganesson ($Z=118$), and beyond for all
hypothetical superheavy elements up to $Z\le137$. 
Although superseded by the Dirac equation and modern quantum electrodynamics, 
the semiclassical approach offers a historically and pedagogically valuable perspective. 
Using the Sommerfeld fine structure formula and computer algebra methods, 
we demonstrate the appearance of self-intersecting orbits in super strong to ultra strong
Coulomb fields, beginning with Oganesson, and so on for all hypothetical elements when $Z\le137$. 
These orbits can be classified by their `winding' numbers, providing a simple topological
description of Coulomb field strength in this framework. 
Our results highlight a conceptual bridge between early quantum theory and modern
superheavy element physics.
\end{abstract}

\dedicatory{\begin{center}
{\scriptsize{Dedicated to the memory of Professor Alladi Ramakrishnan}}
\end{center}}

\maketitle

\section{Transuranics, Fine Structure Formula, and Orbits\/}

Transuranic elements are chemical elements with atomic numbers greater than $%
92$ (the atomic number of uranium). All of them are unstable and
radioactive, and most must be synthesized in a laboratory \cite{Kragh18}.
The high positive charge of the nucleus creates an extremely strong electric
field that affects the behavior of all orbiting electrons. For the innermost
electrons, the electrostatic attraction is so strong that their velocities
approach the speed of light.

The Bohr-Sommerfeld model was an early attempt to explain atomic structure
by incorporating elliptical orbits and special relativity into the Bohr
model \cite{Barleyetal2025, Kragh2012, KraghBohr}. However, its simplified,
semiclassical approach has fundamental limitations that prevent it from
accurately describing transuranic elements. Nonetheless, we would like to
explore this ``missing opportunity" for highly charged superheavy
transuranic one-electron systems.

The original Sommerfeld fine structure formula is given by \cite{Somm1916,
SomAS}: 
\begin{equation}
\frac{E_{n_{r},n_{\theta }}}{mc^{2}}=\left[ 1+\frac{\alpha ^{2}Z^{2}}{\left(
n_{r}+\left( n_{\theta }^{2}-\alpha ^{2}Z^{2}\right) ^{1/2}\right) ^{2}}%
\right] ^{-1/2} \ \ (Z\le137),  \label{BidSomEnd}
\end{equation}%
where $n_{r}\geq 0$ (the {\textit{radial quantum number}}) and $n_{\theta }\geq 1$ (the {%
\textit{azimuthal quantum number}}) are integers. This result made
it possible to explain, for the first time, the fine structure of spectral
lines. (For further details, see \cite{Barleyetal2021, Barleyetal2025,
Elyashevich1985, KraghBohr, Reed} and the references therein.) 

In standard geometrical terms, the orbit equation is%
\begin{equation}
\frac{1}{r}=\frac{1+\epsilon \cos \left( \omega \theta \right) }{a\left(
1-\epsilon ^{2}\right) }.  \label{BidSom9}
\end{equation}%
With the eccentricity $\epsilon $, we have, for $\omega \theta = \phi =0$,
the perihelion distance $r_{\text{min}}=a(1-\epsilon )$, and for $\omega
\theta = \phi =\pi $, the aphelion distance $r_{\text{max}}=a(1+\epsilon )$.

As stated in original works \cite{Biedenharn1983, LaLif2, SomAS}, classical
relativistic Kepler orbits have the form of conic sections as in the
nonrelativistic case, but with a new angular variable $\phi =\omega \theta .$
Thus, for elliptical orbits (bound states), the motion from one perihelion ($%
\phi =0$) to the next ($\phi =2\pi $) requires $\theta =2\pi /\omega \,$,
with a per-revolution shift of $\Delta \theta =2\pi /\omega -2\pi .$%
\footnote{%
In the terminology of classical work \cite{SomAS} where only sufficiently
small $\Delta \theta $ are discussed (Figure~\ref{Figure1}). Throughout this
article we shall keep the original name, `elliptical' curve, although for
certain $\omega$'s a different topological structure may occur such as
orbits with different `winding' numbers and with multiple points of
self-intersections (see, for example, Figures~\ref{Figure2} and \ref{Figure3} below).}

Quantized values of parameters of the electron's elliptical orbits (\ref%
{BidSom9}) are evaluated in \cite{Barleyetal2025} as follows%
\begin{eqnarray}
\omega _{n_{\theta }}n_{\theta } &=&\left( n_{\theta }^{2}-\alpha
^{2}Z^{2}\right) ^{1/2},  \label{BidSom18} \\
\epsilon _{n_{r},n_{\theta }} &=&\sqrt{n_{r}}\cdot \frac{\left( n_{r}+2\sqrt{%
n_{\theta }^{2}-\alpha ^{2}Z^{2}}\right) ^{1/2}}{n_{r}+\sqrt{n_{\theta
}^{2}-\alpha ^{2}Z^{2}}},  \label{BidSom19} \\
a_{n_{r},n_{\theta }} &=&\frac{a_{0}}{Z}\left( n_{r}+\sqrt{n_{\theta
}^{2}-\alpha ^{2}Z^{2}}\right)  \label{BidSom20} \\
&&\times \sqrt{\alpha ^{2}Z^{2}+\left( n_{r}+\sqrt{n_{\theta }^{2}-\alpha
^{2}Z^{2}}\right) ^{2}} ,  \notag
\end{eqnarray}%
where $a_{0}=\hbar ^{2}/(me^{2})$ is the familiar Bohr atomic radius and $\alpha =
e^{2} / (\hbar c)$ is the Sommerfeld fine structure constant. These formulas generalize
the classical circular orbits.\footnote{%
Classical solutions of the relativistic Kepler problem are also discussed in 
\cite[pp.~481--482]{Goldstein} and \cite[pp.~100--102]{LaLif2}.} 
The size and shape of the orbit are fixed in accordance with the azimuthal and radial quantization conditions
-- a discrete family of quantized orbits is selected from the continuous set of all possible solutions \cite{SomAS}.

We found the following concept useful for `winding' numbers. Let us introduce
\begin{equation}\label{WNumber}
 N_{\rm{winding}}:=2 \left(\dfrac{1}{\omega} - 1\right),
\end{equation}
with approximation to the nearest integer, 
in the case of non-closed relativistic orbits under consideration.
It usually coincides with a direct count of the electron's rotation around the focus in one orbital period.
Moreover, one may count only rotations, say from $r_{\rm{min}}$ to $r_{\rm{max}}$, due to the following symmetry: $r(2\pi/\omega-\theta)=r(\theta)$ and then double the result.

The speed of an electron can also be estimated as follows:
\begin{equation}\label{GroundSpeed}
\dfrac{v_{\rm{ground}}}{c}=\alpha Z
\end{equation}
for the ground state, when  $n_{r}=0$ and $n_{\theta}=1$ (see Appendix~A). 
As a result, the corresponding relativistic de Broglie wavelength is given by
\begin{equation}\label{deBroglie}
\lambda_{\textrm{ground}}^{\textrm{dB}} = \dfrac{h}{p}=\dfrac{2\pi\hbar}{mc\alpha Z} \sqrt{1-(\alpha Z)^2}\/.
\end{equation}
This work is a continuation of \cite{Barleyetal2021, Barleyetal2025, BarleySusMathTwo, BarleyRufSusOg}.
Here we present data on semiclassical orbits for the relativistic electron's first excited state, with nonzero angular momentum, for all transuranic elements: $92\le Z \le 137$.

We dedicate this article to the memory of Professor Alladi Ramakrishnan in view of his important contributions to probability theory and stochastic processes, the theory of fundamental particles, special relativity and relativistic quantum mechanics, mathematical sciences and education \cite{AlladiKlauderRao, AlRamaProb, AlRamaParticles, AlRamaDiracMat, AlRamaDiracSpecialR, AlRamaDiary}.

\section{Heavy one- electron Ions: From Uranium through Oganesson to Hypothetical Feynmanium\/}

In this section we present results of computer simulations for all transuranic elements, $92\le Z \le 137\/,$ in the case of the simplest excited state with nonzero angular momentum, namely, $n_{r}=n_{\theta}=1$, on the basis of the `old' Bohr-Sommerfeld relativistic model. Our complementary Mathematica notebook, Transuranics.nb \cite{TransU}, see also \cite{BarleySusMathTwo}, contains all animation details,
allowing the reader to repeat our calculations and explore other cases.

In the notebook introductory section, the general formulas are provided. They are stored in global variables that will be used in all the subsequent sections. For this purpose, allow Mathematica to evaluate all initialization cells. After that you will be able to run each case independently from the others. (Always use the Clear[...] command to remove old data.)

We emphasize that the present consideration is intended as a conceptual and pedagogical exploration, not as a predictive alternative to relativistic quantum mechanics, in a way,
following Mendel{\'e}ev's approach -- getting ``Mach out of Little" \cite{Wheeler1971}.

The orbit parameters under consideration are collected in Tables~\ref{Table1}--\ref{Table9} for the reader's convenience.%
\footnote{
We use the standard abbreviations CW and CCW for clockwise and counterclockwise rotation of the ``rosettes'', respectively.
}  

\subsection{Uranium (U), $Z=92\/$} Uranium is a naturally occurring, dense, silvery-white radioactive element with the atomic number \(92\). 
It is a key component in nuclear energy, as the isotope, U-235, can be enriched to sustain a nuclear chain reaction for power nuclear stations or weapons.
Quantum electrodynamics of the hydrogen-like uranium ion has been studied in \cite{Gum2005, Gum2007} (see also \cite{FleKar1971, Mohretal1998, Oganessian2017, ShabQED, ShabQEDHI, SusRelInt} and references therein).
%
%
We have used Mathematica for the following animations \cite{BarleySusMathTwo, TransU}.

\begin{figure}[hbt!]
\centering
\includegraphics[width=0.557\textwidth]{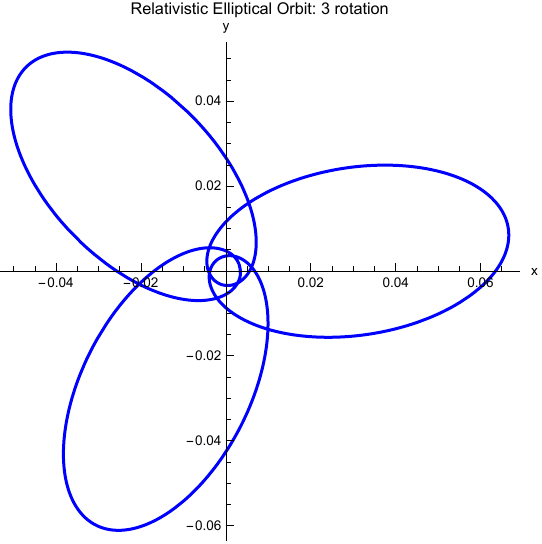}
\caption{Kepler's Classical Elliptical Curve Counterclockwise Motion in Relativistic Uranium Hydrogen-like
Ion U$^{91+}\/.$ The `winding' number is one.}
\label{Figure1}
\end{figure}

In the Bohr-Sommerfeld atomic model under consideration, even for the strong electric field of the Uranium nucleus, 
a single electron stays on classical relativistic elliptical orbits with advancing perihelion, 
in a form of open ``rosettes" \cite{Biedenharn1983, LaLif2, SomAS} (see Figure~\ref{Figure1} with the standard orbit `winding' number one).  
(Table~\ref{Table1}, column~2)
%

%
%

\subsection{Neptunium (Np), $Z=93\/$} Neptunium is a radioactive, silvery-gray, synthetic element in the actinide series, named after the planet Neptune. It was the first transuranium element to be artificially produced, discovered in 1940 by Edwin McMillan and Philip Abelson. 
(Table~\ref{Table1}, column~3)

\subsection{Plutonium (Pu), $Z=94\/$} Plutonium is a radioactive, silvery-gray metallic element with the atomic number \(94\). It is an important element in nuclear weapons and reactors, but is mostly produced in nuclear reactors from uranium, though trace amounts can be found naturally.
(Table~\ref{Table1}, column~4)

\subsection{Americium (Am), $Z=95\/$} Americium is a synthetic, radioactive, silvery-white metal with the atomic number \(95\) and the symbol Am. It is a transuranic element, belonging to the actinide series, and is produced as a byproduct of nuclear reactors \cite{Myasoedov1985}.
(Table~\ref{Table1}, column~5)

\subsection{Curium (Cm), $Z=96\/$} Curium is a radioactive, synthetic element in the actinide series, with the atomic number \(96\), that is named after Marie and Pierre Curie. It is a hard, silvery-white metal that is primarily used in scientific research, for example, as a heat and alpha particle source \cite{Myasoedov1985}.
(Table~\ref{Table1}, column~6)

%
\begin{table}[t]
\begin{tabular}{|l|l|l|l|l|l|}
\hline
$\text{Parameters \ vs} \ \ Z$ & U$^{91+}$ Ion  & Np$^{92+}$ Ion & Pu$^{93+}$ Ion    & Am$^{94+}$ Ion  & Cm$^{95+}$ Ion\\ 
\hline
$\omega$                       & $0.741\ 135$   & $0.734\ 458$   & $0.727\ 648$      & $0.720\ 699$    & $0.713\ 607$\\ 
\hline
$\epsilon$                     & $0.904\ 882$   & $0.905\ 919$   & $0.906\ 978$      & $0.908\ 06$     & $0.9091\ 65$\\ 
\hline
$a/a_{0}$                      & $0.035\ 3163$  & $0.034\ 7358$  & $0.034\ 1641$     & $0.033\ 6007$   & $0.033\ 0454$\\ 
\hline
$r_{\text{min}}$               & $0.003\ 35921$ & $0.003\ 26798$ & $0.003\ 17802$    & $0.003\ 08926$  & $0.003\ 00167$\\ 
\hline
$r_{\text{max}}$               & $0.067\ 2735$  & $0.0662\ 037$  & $0.065\ 1502$     & $0.064\ 1122$   & $0.063\ 0891$\\ 
\hline
$\Delta\theta$                 & $2.194\ 61$    & $2.271\ 67$    & $2.351\ 74$       & $2.435$         & $2.521\ 64$\\ 
\hline
$v_{\rm{ground}}/c$            & $0.672$        & $0.679$        & $0.686$           & $0.693$         & $0.701$\\ 
\hline
$E/mc^2$                       & $0.933\ 042$   & $0.931\ 251$   & $0.929\ 421$      & $0.927\ 55$     & $0.925\ 637$\\ 
\hline
`Winding' numbers              & $0.699$ CCW    & $0.723$ CCW    & $0.749$ CCW       & $0.775$ CCW     & $0.802$ CCW\\ 
\hline
\end{tabular}
\centering
\caption{Orbit parameters for U$^{91+}$, Np$^{92+}$, Pu$^{93+}$, Am$^{94+}$, and Cm$^{95+}$ ions.}
\label{Table1}
\end{table}
%

%
\subsection{Berkelium (Bk), $Z=97\/$} Berkelium is a synthetic, radioactive transuranic element with the atomic number \(97\). It is a soft, silvery-white metal named after Berkeley, California, where it was discovered in 1949. Its primary use is in scientific research to synthesize heavier elements. (Table~\ref{Table2}, column~2)

\subsection{Californium (Cf), $Z=98\/$} Californium is a synthetic, radioactive chemical element with the atomic number \(98\). It is an actinide element that is produced by bombarding curium with alpha particles and has found practical uses in areas like nuclear reactor start-up, cancer therapy, and neutron analysis due to its potent neutron-emitting properties. It was discovered in 1950 and named after the state of California and the university where it was first synthesized.  
(Table~\ref{Table2}, column~3)

\subsection{Einsteinium (Es), $Z=99\/$} Einsteinium is a synthetic, radioactive element with the atomic number \(99\), named after Albert Einstein. 
(Table~\ref{Table2}, column~4)

\subsection{Fermium (Fm), $Z=100\/$} Fermium is a synthetic, radioactive actinide element with the atomic number \(100\). 
This transuranic element cannot be found in nature and must be artificially created, primarily for scientific research in nuclear physics.
Fermium was discovered in 1953 in the debris of the first hydrogen bomb test. It was named after the physicist Enrico Fermi.  
(Table~\ref{Table2}, column~5)

\subsection{Mendelevium (Md), $Z=101\/$} Mendelevium is a synthetic, radioactive, transuranic actinide element with the atomic number \(101\). It was named after Dmitri~I.~Mendel{\'e}ev, the creator of the periodic table, and was first discovered in 1955 by bombarding Einsteinium with alpha particles.
(Table~\ref{Table2}, column~6)

\begin{table}[t]
\begin{tabular}{|l|l|l|l|l|l|}
\hline
$\text{Parameters \ vs} \ \ Z$ & Bk$^{96+}$ Ion & Cf$^{97+}$ Ion & Es$^{98+}$ Ion    & Fm$^{99+}$ Ion  & Md$^{100+}$ Ion\\ 
\hline
$\omega$                       & $0.706\ 37$   & $0.698\ 981$    & $0.691\ 436$      & $0.683\ 73$     & $0.675\ 857$\\ 
\hline
$\epsilon$                     & $0.910\ 295$   & $0.911\ 45$    & $0.912\ 63$       & $0.913\ 837$    & $0.915\ 072$\\ 
\hline
$a/a_{0}$                      & $0.032\ 4977$  & $0.031\ 9574$  & $0.031\ 4241$     & $0.030\ 8975$   & $0.030\ 3773$\\ 
\hline
$r_{\text{min}}$               & $0.002\ 91521$ & $0.002\ 82984$ & $0.002\ 74552$    & $0.002\ 66221$  & $0.002\ 57988$\\ 
\hline
$r_{\text{max}}$               & $0.062\ 0802$  & $0.061\ 085$   & $0.060\ 1027$     & $0.059\ 1328$   & $0.058\ 1747$\\ 
\hline
$\Delta\theta$                 & $2.611\ 85$    & $2.705\ 88$    & $2.803\ 97$       & $2.906\ 39$     & $3.013\ 43$\\ 
\hline
$v_{\rm{ground}}/c$            & $0.708$        & $0.715$        & $0.723$           & $0.730$         & $0.737$\\ 
\hline
$E/mc^2$                       & $0.923\ 68$   & $0.921\ 678$    & $0.919\ 629$      & $0.917\ 532$    & $0.915\ 384$\\ 
\hline
`Winding' numbers              & $0.831$ CCW    & $0.861$ CCW    & $0.893$ CCW       & $0.925$ CCW     & $0.959$ CCW\\ 
\hline
\end{tabular}
\centering
\caption{Orbit parameters for Bk$^{96+}$, Cf$^{97+}$, Es$^{98+}$, Fm$^{99+}$, and Md$^{100+}$ ions.}
\label{Table2}
\end{table}

\subsection{Nobelium (No), $Z=102\/$} Nobelium is a synthetic, radioactive actinide element with the atomic number \(102\). It is a highly radioactive metal that does not occur naturally and is produced in labs for scientific research, such as studying the properties of super-heavy elements.
(Table~\ref{Table3}, column~2)

\subsection{Lawrencium (Lr), $Z=103\/$} Lawrencium is a synthetic, radioactive element with the atomic number \(103\). It is an actinide, named after Ernest O. Lawrence, and is created by bombarding Californium with Boron ions.
(Table~\ref{Table3}, column~3)

\subsection{Rutherfordium (Rf), $Z=104\/$} Rutherfordium is a synthetic, highly radioactive element with the atomic number \(104\). It is named after Ernest Rutherford, the father of nuclear physics, and is used in scientific research.
(Table~\ref{Table3}, column~4)

\subsection{Dubnium (Db), $Z=105\/$} Dubnium is a synthetic, highly radioactive chemical element with the atomic number \(105\). As a transition metal, it is expected to be a solid metal at room temperature, though little is known about its physical properties due to its short half-life and lack of practical use in large quantities. It is named after the city of Dubna, Moscow oblast (province), Russia, where it was first synthesized in 1967.  
(Table~\ref{Table3}, column~5)

\subsection{Seaborgium (Sg), $Z=106\/$} Seaborgium is a synthetic, radioactive chemical element with the atomic number \(106\), named after nuclear chemist Glenn T. Seaborg. It is a man-made element, produced in particle accelerators.
(Table~\ref{Table3}, column~6)

\begin{table}[t]
\begin{tabular}{|l|l|l|l|l|l|}
\hline
$\text{Parameters \ vs} \ \ Z$ & No$^{101+}$ Ion & Lr$^{102+}$ Ion & Rf$^{103+}$ Ion   & Db$^{104+}$ Ion  & Sg$^{105+}$ Ion\\ 
\hline
$\omega$                       & $0.667\ 812$    & $0.659\ 588$    & $0.651\ 178$      & $0.642\ 576$     & $0.633\ 773$\\ 
\hline
$\epsilon$                     & $0.916\ 335$    & $0.911\ 45$     & $0.918\ 951$      & $0.920\ 306$    & $0.921\ 693$\\ 
\hline
$a/a_{0}$                      & $0.029\ 8631$   & $0.029\ 3547$   & $0.028\ 8518$     & $0.028\ 354$    & $0.027\ 861$\\ 
\hline
$r_{\text{min}}$               & $0.002\ 49849$  & $0.002\ 41801$  & $0.002\ 33841$    & $0.002\ 25965$  & $0.002\ 18171$\\ 
\hline
$r_{\text{max}}$               & $0.057\ 2278$   & $0.056\ 2915$    & $0.055\ 3652$    & $0.054\ 4483$   & $0.053\ 5403$\\ 
\hline
$\Delta\theta$                 & $3.125\ 43$     & $3.242\ 74$     & $3.365\ 76$       & $3.494\ 94$     & $3.630\ 76$\\ 
\hline
$v_{\rm{ground}}/c$            & $0.745$         & $0.752$         & $0.759$           & $0.766$         & $0.774$\\ 
\hline
$E/mc^2$                       & $0.913\ 185$     & $0.910\ 93$   & $0.908\ 619$       & $0.906\ 249$    & $0.903\ 818$\\ 
\hline
`Winding' numbers              & $0.995$ CW     & $1.032$ CCW      & $1.071$ CCW       & $1.112$ CCW     & $1.156$ CCW\\ 
\hline
\end{tabular}
\centering
\caption{Orbit parameters for No$^{101+}$, Lr$^{102+}$, Rf$^{103+}$, Db$^{104+}$, and Sg$^{105+}$ ions.}
\label{Table3}
\end{table}

\subsection{Bohrium (Bh), $Z=107\/$} Bohrium is a synthetic, radioactive element with the atomic number \(107\), named after the physicist Niels Bohr, the father of quantum physics. It is produced in particle accelerators by bombarding other nuclei.
(Table~\ref{Table4}, column~2)

\subsection{Hassium (Hs), $Z=108\/$} Hassium is a synthetic, radioactive transition metal with the atomic number \(108\). It is named after the German Land (state) of Hessen, where it was first discovered in 1984 by a team led by Peter Armbruster and Gottfried M\"{u}nzenberg.
(Table~\ref{Table4}, column~3)

\subsection{Meitnerium (Mt), $Z=109\/$} Meitnerium is a synthetic, radioactive element with the atomic number \(109\), named after female physicist Lise Meitner. Created in Germany in 1982, it is a transuranic metal with an unknown appearance and no practical uses beyond scientific research.
(Table~\ref{Table4}, column~4)

\subsection{Darmstadtium (Ds), $Z=110\/$} Darmstadtium is a synthetic, highly radioactive element with the atomic number \(110\), discovered in 1994 at the GSI Helmholtz Centre for Heavy Ion Research in Germany. Named after the city of Darmstadt, it is a transition metal that is predicted to have metallic properties but is only created for research purposes due to its instability and extremely short half-life.  
(Table~\ref{Table4}, column~5)

\subsection{Roentgenium (Rg), $Z=111\/$} Roentgenium is a synthetic, radioactive chemical element with the atomic number \(111\) first synthesized in 1994 at the GSI Helmholtz Centre for Heavy Ion Research in Germany. Named after German physicist Wilhelm Conrad R{\"o}ntgen, the discoverer of X-rays.  
(Table~\ref{Table4}, column~6)
%

%
\begin{table}[t]
\begin{tabular}{|l|l|l|l|l|l|}
\hline
$\text{Parameters \ vs} \ \ Z$ & Bh$^{106+}$ Ion & Hs$^{107+}$ Ion & Mt$^{108+}$ Ion   & Ds$^{109+}$ Ion  & Rg$^{110+}$ Ion\\ 
\hline
$\omega$                       & $0.624\ 76$     & $0.615\ 529$    & $0.606\ 07$       & $0.596\ 371$     & $0.586\ 421$\\ 
\hline
$\epsilon$                     & $0.923\ 115$    & $0.924\ 572$    & $0.926\ 066$      & $0.927\ 598$    & $0.929\ 171$\\ 
\hline
$a/a_{0}$                      & $0.027\ 3725$   & $0.026\ 8883$   & $0.026\ 408$      & $0.025\ 9312$    & $0.025\ 4577$\\ 
\hline
$r_{\text{min}}$               & $0.002\ 10454$  & $0.002\ 02814$  & $0.001\ 95245$    & $0.001\ 87747$  & $0.001\ 80315$\\ 
\hline
$r_{\text{max}}$               & $0.052\ 6405$   & $0.051\ 7485$   & $0.050\ 8635$     & $0.049\ 985$   & $0.049\ 1123$\\ 
\hline
$\Delta\theta$                 & $3.773\ 77$     & $3.924\ 59$     & $4.08\ 391$       & $4.252\ 51$     & $4.431\ 27$\\ 
\hline
$v_{\rm{ground}}/c$            & $0.781$         & $0.788$         & $0.796$           & $0.803$         & $0.81$\\ 
\hline
$E/mc^2$                       & $0.901\ 321$    & $0.898\ 757$    & $0.896\ 122$      & $0.893\ 412$    & $0.890\ 624$\\ 
\hline
`Winding' numbers              & $1.201$ CCW     & $1.249$ CCW     & $1.3$ CCW         & $1.354$ CCW     & $1.411$ CCW\\ 
\hline
\end{tabular}
\centering
\caption{Orbit parameters for Bh$^{106+}$, Hs$^{107+}$, Mt$^{108+}$, Ds$^{109+}$, and Rg$^{110+}$ ions.}
\label{Table4}
\end{table}

\subsection{Copernicium (Cn), $Z=112\/$} Copernicium is a synthetic, radioactive chemical element with the atomic number \(112\), discovered in 1996 and named after astronomer Nicolaus Copernicus. 
(Table~\ref{Table5}, column~2) 

\subsection{Nihonium (Nh), $Z=113\/$} Nihonium is a synthetic, radioactive chemical element with the atomic number \(113\). It was discovered in 2004 by scientists at the Japanese research institute RIKEN and named ``Nihonium" in honor of Japan (Nihon).
(Table~\ref{Table5}, column~3) 

\subsection{Flerovium (Fl), $Z=114\/$} Flerovium is a synthetic, radioactive element with the atomic number \(114\), discovered in 1999 by scientists at the Joint Institute for Nuclear Research in Russia and the Lawrence Livermore National Laboratory in the United States. It is a superheavy metal, named after the Flerov Laboratory of Nuclear Reactions and its founder, Soviet physicist Georgy N.~Flerov.
(Table~\ref{Table5}, column~4)

\subsection{Moscovium (Mc), $Z=115\/$} Moscovium is a synthetic, radioactive element with the atomic number \(115\), discovered in 2003 by scientists at the Joint Institute for Nuclear Research and Lawrence Livermore National Laboratory. Named after the Moscow Oblast region of Russia, where the Joint Institute for Nuclear Research is located. 
(Table~\ref{Table5}, column~5)

\subsection{Livermorium (Lv), $Z=116\/$} Livermorium is a synthetic, radioactive element with the atomic number 116, named to honor Lawrence Livermore National Laboratory. 
(Table~\ref{Table5}, column~6)

\begin{table}[t]
\begin{tabular}{|l|l|l|l|l|l|}
\hline
$\text{Parameters \ vs} \ \ Z$ & Cn$^{111+}$ Ion & Nh$^{112+}$ Ion & Fl$^{113+}$ Ion   & Mc$^{114+}$ Ion  & Lv$^{115+}$ Ion\\ 
\hline
$\omega$                       & $0.576\ 207$    & $0.565\ 715$    & $0.554\ 928$      & $0.543\ 83$      & $0.532\ 4$\\ 
\hline
$\epsilon$                     & $0.930\ 786$    & $0.932\ 444$    & $0.934\ 149$      & $0.935\ 902$     & $0.937\ 706$\\ 
\hline
$a/a_{0}$                      & $0.024\ 9872$   & $0.024\ 5192$   & $0.024\ 0534$     & $0.023\ 5894$    & $0.023\ 1268$\\ 
\hline
$r_{\text{min}}$               & $0.001\ 72947$  & $0.001\ 65641$  & $0.001\ 58394$    & $0.001\ 51203$   & $0.001\ 44065$\\ 
\hline
$r_{\text{max}}$               & $0.048\ 2449$   & $0.047\ 3819$   & $0.046\ 5228$     & $0.045\ 6667$    & $0.044\ 8129$\\ 
\hline
$\Delta\theta$                 & $4.621\ 2$      & $4.823\ 44$     & $5.039\ 33$       & $5.270\ 4$       & $5.518\ 44$\\ 
\hline
$v_{\rm{ground}}/c$            & $0.818$         & $0.825$         & $0.832$           & $0.839$          & $0.847$\\ 
\hline
$E/mc^2$                       & $0.887\ 752$    & $0.884\ 792$    & $0.881\ 739$      & $0.878\ 587$     & $0.875\ 329$\\ 
\hline
`Winding' numbers              & $1.471$ CCW     & $1.535$ CCW     & $1.604$ CCW       & $1.678$ CCW      & $1.756$ CW\\ 
\hline
\end{tabular}
\centering
\caption{Orbit parameters for Cn$^{111+}$, Nh$^{112+}$, Fl$^{113+}$, Mc$^{114+}$, and Lv$^{115+}$ ions.}
\label{Table5}
\end{table}

\subsection{Tennessine (Ts), $Z=117\/$} Tennessine is the synthetic, superheavy radioactive element with the atomic number \(117\). It was officially named to honor the contribution of the US state of Tennessee to the element's discovery, which involved researchers from the Oak Ridge National Laboratory (ORNL), Vanderbilt University, and the University of Tennessee. The element was first detected in 2010 at Russia's Joint Institute for Nuclear Research (JINR).
(Table~\ref{Table6}, column~2)

\subsection{Oganesson (Og), $Z=118\/$} 
\begin{figure}[tbh]
\centering
\includegraphics[width=0.557\textwidth]{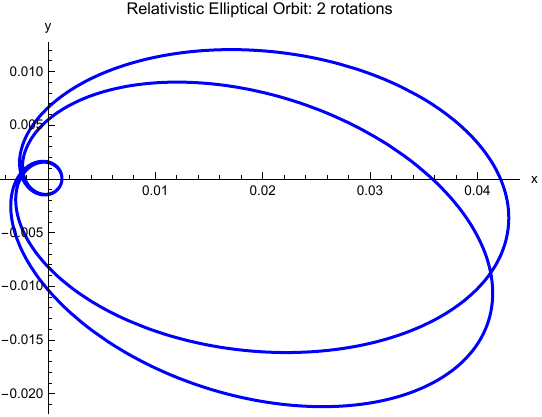}
\caption{Kepler's `Elliptical' Curve Clockwise Motion with Self-intersection in
Hypothetical Oganesson Hydrogen-like Ion Og$^{117+}\/.$ The `winding' number is two.}
\label{Figure2}
\end{figure}

\begin{figure}[tbh]
\centering
\includegraphics[width=0.557\textwidth]{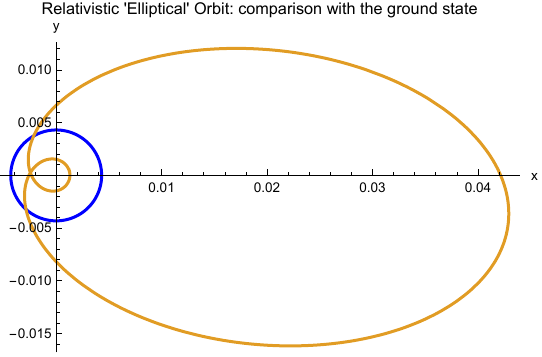}
\caption{Kepler's `Elliptical' Orbit with Self-intersection, $n_{\theta}=n_{r}=1\/,$ in
Hypothetical Oganesson Hydrogen-like Ion Og$^{117+}\/$ in comparison with 
the ground state, when $n_{\theta}=1$ and $n_{r}=0\/.$ The `winding' number is two.}
\label{Figure3}
\end{figure}

Oganesson is a synthetic element with the atomic number \(118\) -- the highest atomic number of any element. Discovered in 2006, Dubna, Moscow region, and  named after physicist Yuri Ts.~Oganessian, a pioneer in the field of superheavy elements. 
(Table~\ref{Table6}, column~3) On Figures~\ref{Figure2} and \ref{Figure3} one can observe, for the first time, an extra loop -- the so-called ``chemical orbit'' or ``double necklace'' in the terminology of the American cosmologist John Archibald Wheeler \cite{CeulemansThyssen2018, Powers1971, Wheeler1971}.
(See also \cite%
{Chap2018, Colloquium2019, Kaygorodovetal2021, Kragh18, Oganessian2006, Oganessian2012} and references therein). 

The radius $R$ of a nucleus can be estimated as follows 
$$
R=R_0A^{1/3}, \qquad R_0 \approx 1.2 \times 10^{-13} {\textrm{cm}} .
$$ 
For oganesson, the mass number $A=294$ and $R_{\textrm{Og}} \approx 0.798 \times 10^{-12} {\textrm{cm}}\/.$
In Bohr's atomic units: $R_{\textrm{Og}}/a_0 \approx 1.51 \times 10^{-4}$ and $r_{\text{min}}\approx 8.61 \times R_{\textrm{Og}}\/.$
For the ground state by (\ref{deBroglie}), one gets
\begin{equation*}\label{deBroglieGround}
\lambda_{\textrm{ground}}^{\textrm{dB}} \approx 1.43 \times 10^{-10} {\textrm{cm}}\/,
\end{equation*}
or $\lambda_{\textrm{ground}}^{\textrm{dB}} \approx 179 \times R_{\textrm{Og}}\/.$

%
%

%
%
%
%

\subsection{Ununennium (Uue), $Z=119\/$} Hypothetical element E119 with the atomic number \(119\) and a temporary symbol Uue. 
(Table~\ref{Table6}, column~4)

\subsection{Unbinilium (Ubn), $Z=120\/$} Hypothetical element E120 with the atomic number \(120\) and a temporary symbol Ube.
(Table~\ref{Table6}, column~5)

\subsection{Unbiunium (Ubu), $Z=121\/$} Hypothetical element E121 with the atomic number \(121\) and a temporary  symbol Ubu.
(Table~\ref{Table6}, column~6)

\begin{table}[t]
\begin{tabular}{|l|l|l|l|l|l|}
\hline
$\text{Parameters \ vs} \ \ Z$ & Ts$^{116+}$ Ion & Og$^{117+}$ Ion & Uue$^{118+}$ Ion  & Ubn$^{119+}$ Ion & Ubu$^{120+}$ Ion\\ 
\hline
$\omega$                       & $0.520\ 617$    & $0.508\ 457$    & $0.495\ 891$      & $0.482\ 888$      & $0.469\ 411$\\ 
\hline
$\epsilon$                     & $0.939\ 564$    & $0.941\ 479$    & $0.943\ 455$      & $0.945\ 494$     & $0.947\ 601$\\ 
\hline
$a/a_{0}$                      & $0.022\ 6652$   & $0.0222\ 041$   & $0.021\ 7429$     & $0.021\ 2812$    & $0.020\ 183$\\ 
\hline
$r_{\text{min}}$               & $0.001\ 36978$  & $0.001\ 2994$   & $0.001\ 22946$    & $0.001\ 15996$   & $0.001\ 09085$\\ 
\hline
$r_{\text{max}}$               & $0.043\ 9606$   & $0.043\ 1087$   & $0.042\ 2564$     & $0.041\ 4024$    & $0.040\ 5457$\\ 
\hline
$\Delta\theta$                 & $5.785\ 54$      & $6.074\ 18$    & $6.387\ 32$       & $6.728\ 51$       & $7.102\ 06$\\ 
\hline
$v_{\rm{ground}}/c$            & $0.854$         & $0.861$         & $0.869$           & $0.875\ 912$          & $0.883$\\ 
\hline
$E/mc^2$                       & $0.871\ 957$    & $0.868\ 463$    & $0.864\ 838$      & $0.861\ 071$     & $0.857\ 15$\\ 
\hline
`Winding' numbers              & $1.842$ CW     & $1.933$ CW       & $2.033\ 15$ CCW   & $2.142$ CCW      & $2.261$ CCW\\ 
\hline
\end{tabular}
\centering
\caption{Orbit parameters for Ts$^{116+}$, Og$^{117+}$, Uue$^{118+}$, Ubn$^{119+}$, and Ubu$^{120+}$ ions.}
\label{Table6}
\end{table}

\subsection{Unbibium (Ubb), $Z=122\/$} Hypothetical element E122 with the atomic number \(122\) and a temporary symbol Ubb.
(Table~\ref{Table7}, column~2)

\subsection{Unbitrium (Ubt), $Z=123\/$} Hypothetical element E123 with the atomic number \(123\) and a temporary symbol Ubt.
(Table~\ref{Table7}, column~3)

\subsection{Unbiquadium (Ubq), $Z=124\/$} Hypothetical element E124 with the atomic number \(124\) and a temporary symbol Ubq.
(Table~\ref{Table7}, column~4)

\subsection{Unbipentium (Ubp), $Z=125\/$} Hypothetical element E125 with the atomic number \(125\) and a temporary symbol Ubp.
(Table~\ref{Table7}, column~5)
%

%
\begin{figure}[hbt!]
\centering
\includegraphics[width=0.375\textwidth]{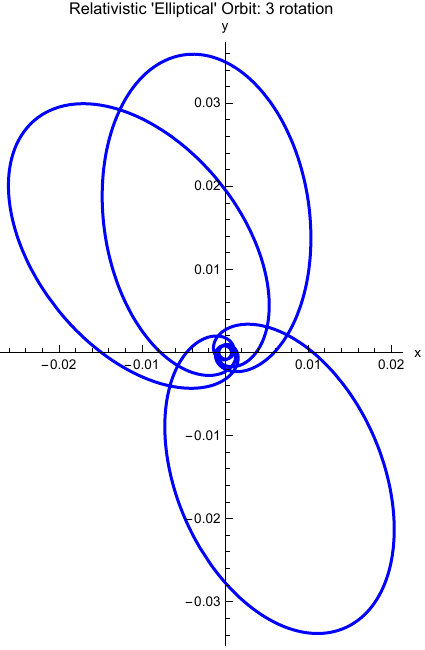}
\caption{Kepler's `Elliptical' Curve Counterclockwise Motion in Hypothetical Unbihexium Relativistic Hydrogen-like
Ion Ubh$^{125+}\/.$ The `winding' number is three.}
\label{Figure4}
\end{figure}
%

\subsection{Unbihexium (Ubh), $Z=126\/$} Hypothetical element E126 with the atomic number \(126\) and a temporary symbol Ubh.
(Table~\ref{Table7}, column~6, Figure~4) 

%
\begin{table}[t]
\begin{tabular}{|l|l|l|l|l|l|}
\hline
$\text{Parameters \ vs} \ \ Z$ & Ubb$^{121+}$ Ion & Ubt$^{122+}$ Ion  & Ubq$^{123+}$ Ion  & Ubp$^{124+}$ Ion & Ubh$^{125+}$ Ion\\ 
\hline
$\omega$                       & $0.455\ 419$     & $0.440\ 863$      & $0.425\ 684$      & $0.409\ 814$      & $0.393\ 169$\\ 
\hline
$\epsilon$                     & $0.949\ 782$     & $0.952\ 041$      & $0.954\ 384$      & $0.956\ 818$     & $0.959\ 352$\\ 
\hline
$a/a_{0}$                      & $0.020\ 3534$    & $0.019\ 8858$     & $0.019\ 4146$     & $0.0189\ 386$    & $0.018\ 4565$\\ 
\hline
$r_{\text{min}}$               & $0.001\ 02211$   & $0.0009\ 53712$   & $0.0008\ 85619$   & $0.0008\ 178$   & $0.0007\ 50218$\\ 
\hline
$r_{\text{max}}$               & $0.039\ 6847$    & $0.038\ 818$      & $0.037\ 9436$     & $0.037\ 0594$    & $0.036\ 1628$\\ 
\hline
$\Delta\theta$                 & $7.5133$         & $7.968\ 82$       & $8.477\ 01$       & $9.048\ 61$       & $9.697\ 71$\\ 
\hline
$v_{\rm{ground}}/c$            & $0.891$          & $0.898$           & $0.905$           & $0.912$          & $0.920$\\ 
\hline
$E/mc^2$                       & $0.853\ 059$     & $0.848\ 782$      & $0.8443$          & $0.839\ 587$     & $0.834\ 616$\\ 
\hline
`Winding' numbers              & $2.392$ CCW      & $2.536$ CCW       & $2.698$ CCW       & $2.880$ CCW      & $3.087$ CCW\\ 
\hline
\end{tabular}
\centering
\caption{Orbit parameters for Ubb$^{121+}$, Ubt$^{122+}$, Ubq$^{123+}$, Ubp$^{124+}$, and Ubh$^{125+}$ ions.}
\label{Table7}
\end{table}
%

%
\subsection{Unbiseptium (Ubs), $Z=127\/$} Hypothetical element E127 with the atomic number \(127\) and a temporary symbol Ubh.
(Table~\ref{Table8}, column~2)

\subsection{Unbiocitium (Ubo), $Z=128\/$} hypothetical element E128 with the atomic number \(128\) and a temporary symbol Ubo.
(Table~\ref{Table8}, column~3) 

\subsection{Unbiennium (Ube), $Z=129\/$} Hypothetical element E129 with the atomic number \(129\) and a temporary symbol Ube.
(Table~\ref{Table8}, column~4) 

\subsection{Untriunium (Utu), $Z=131\/$} Hypothetical element E131 with the atomic number \(131\) and a temporary symbol Utu.
(Table~\ref{Table8}, column~5) 

\subsection{Untribium (Utb), $Z=132\/$} Hypothetical element E132 with the atomic number \(132\) and a temporary symbol Utb.
(Table~\ref{Table8}, column~6)

%
\begin{table}[t]
\begin{tabular}{|l|l|l|l|l|l|}
\hline
$\text{Parameters \ vs} \ \ Z$ & Ubs$^{126+}$ Ion & Ubo$^{127+}$ Ion  & Ube$^{128+}$ Ion  & Utn$^{129+}$ Ion & Utu$^{130+}$ Ion\\ 
\hline
$\omega$                       & $0.375\ 645$     & $0.357\ 113$      & $0.337\ 408$      & $0.316\ 31$      & $0.293\ 519$\\ 
\hline
$\epsilon$                     & $0.961\ 995$     & $0.964\ 757$      & $0.967\ 653$      & $0.970\ 699$     & $0.973\ 915$\\ 
\hline
$a/a_{0}$                      & $0.0179\ 668$    & $0.0174\ 674$     & $0.016\ 9559$     & $0.016\ 4289$    & $0.015\ 8819$\\ 
\hline
$r_{\text{min}}$               & $0.0006\ 82833$  & $0.0006\ 15602$   & $0.0005\ 48474$   & $0.0004\ 81392$  & $0.0004\ 14288$\\ 
\hline
$r_{\text{max}}$               & $0.035\ 2508$    & $0.034\ 3193$     & $0.033\ 3634$     & $0.032\ 3764$    & $0.031\ 3496$\\ 
\hline
$\Delta\theta$                 & $10.44\ 32$      & $11.31\ 12$       & $12.33\ 87$       & $13.58\ 08$      & $15.12\ 32$\\ 
\hline
$v_{\rm{ground}}/c$            & $0.927$          & $0.934$           & $0.942$           & $0.949$          & $0.956$\\ 
\hline
$E/mc^2$                       & $0.829\ 351$     & $0.823\ 745$      & $0.817\ 743$      & $0.811268$       & $0.804\ 214$\\ 
\hline
`Winding' numbers              & $3.324$ CCW      & $3.60$ CW         & $3.927$ CW        & $4.323$ CCW      & $4.814$ CCW\\ 
\hline
\end{tabular}
\centering
\caption{Orbit parameters for 
Ubs$^{126+}$, Ubo$^{127+}$, Ube$^{128+}$, Utn$^{129+}$, and Utu$^{130+}$ ions.}
\label{Table8}
\end{table}
%

\subsection{Untritrium (Utt), $Z=133\/$} Hypothetical E133 element with the atomic number \(133\) and a temporary symbol Utt.
(Table~\ref{Table9}, column~2)

%
\begin{figure}[hbt!]
\centering
\includegraphics[width=0.557\textwidth]{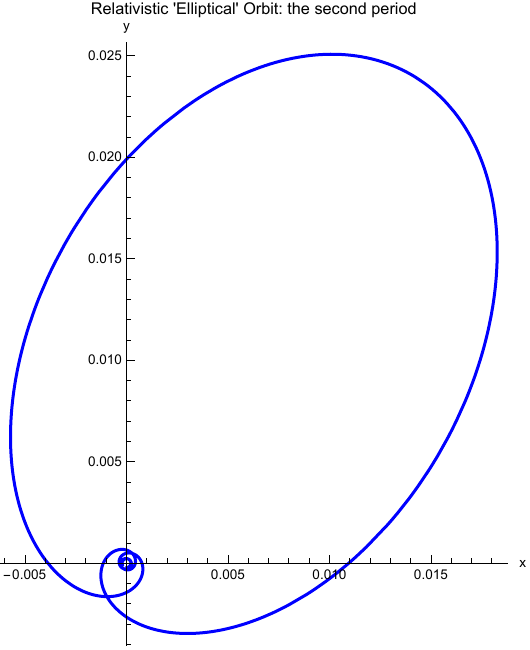}
\caption{Kepler's `Elliptical' Curve Clockwise Motion in Hypothetical Untriquadium Relativistic Hydrogen-like
Ion Utq$^{133+}\/.$ The `winding' number is seven.}
\label{Figure5}
\end{figure}
%

\subsection{ Untriquadium (Utq), $Z=134\/$} Hypothetical E134 element with the atomic number \(134\) and a temporary symbol Utq.
(Table~\ref{Table9}, column~3, Figure~5)

\subsection{Untripentium (Utp), $Z=135\/$} Hypothetical E135 element with the atomic number \(135\) and a temporary symbol Utp.
(Table~\ref{Table9}, column~4)

\subsection{Untrihexium (Uth), $Z=136\/$} Hypothetical E136 element with the atomic number \(136\) and a temporary symbol Uth.
(Table~\ref{Table9}, column~5)

\subsection{Untriseptium (Uts), $Z=137\/$} Hypothetical element E137 (Feynmanium) with the atomic number \(137\) and a temporary symbol Uts.
It is named after physicist Richard Feynman, who pointed out that the electrons in such a heavy element would need to travel at speeds close to the speed of light, leading to theoretical questions about the stability of such an atom.
(Table~\ref{Table9}, column~6)

%
\begin{table}[t]
\begin{tabular}{|l|l|l|l|l|l|}
\hline
$\text{Parameters \ vs} \ \ Z$ & Utt$^{132+}$ Ion & Utq$^{133+}$ Ion  & Utp$^{134+}$ Ion  & Uth$^{135+}$ Ion  & Uts$^{136+}$ Ion\\ 
\hline
$\omega$                       & $0.240\ 908$     & $0.209\ 329$      & $0.171\ 738$      & $0.122\ 731$      & $0.022\ 9203$\\ 
\hline
$\epsilon$                     & $0.980\ 974$     & $0.984\ 905$      & $0.989\ 201$      & $0.994\ 007$      & $0.999\ 749$\\ 
\hline
$a/a_{0}$                      & $0.014\ 6985$    & $0.014\ 0355$     & $0.013\ 287$      & $0.012\ 3706$     & $0.010\ 6797$\\ 
\hline
$r_{\text{min}}$               & $0.0002\ 79652$  & $0.0002\ 11864$   & $0.0001\ 4349$    & $0.00007\ 41347$  & $2.681\ 26 \times 10^{-6}$\\ 
\hline
$r_{\text{max}}$               & $0.029\ 1173$    & $0.027\ 8591$     & $0.026\ 4305$     & $0.024\ 667$      & $0.0213\ 566$\\ 
\hline
$\Delta\theta$                 & $19.798$         & $23.73\ 26$       & $30.30\ 26$       & $44.91\ 15$       & $267.8\ 49$\\ 
\hline
$v_{\rm{ground}}/c$            & $0.971$          & $0.978\ 02$       & $0.985$           & $0.992\ 701$      & $1$\\ 
\hline
$E/mc^2$                       & $0.7876\ 89$     & $0.777602$        & $0.765\ 421$      & $0.749\ 243$      & $0.715\ 164$\\ 
\hline
`Winding' numbers              & $6.302$ CCW      & $7.554$ CW        & $9.645\ 62$ CW    & $14.296$ CCW      & $85.259$ CW\\
\hline
\end{tabular}
\centering
\caption{Orbit parameters for
Utt$^{132+}$, Utq$^{133+}$, Utp$^{134+}$, Uth$^{135+}$, and Uts$^{136+}$ ions.}
\label{Table9}
\end{table}
%

%
{\scshape{Summary and tentative classification:\/}}
According to our calculations, one may separate the strength of Coulomb fields in transuranic elements -- by increasing `winding' numbers -- as follows:
%
\begin{itemize}
\item Strong: $92\le Z \le 116\/$ (no loops).
\item Super-Strong: $117 \le Z \le 125\/$ (one loop, ``double necklace")
\item Ultra-Strong: $126 \le Z \le 128\/$ (two loops, ``triple necklace")
\item Super-Ultra Strong: $129 \le Z \le 130\/$ (three loops)
\item Ultra-Ultra Strong: $131 \le Z \le 137\/$ (many loops).
\end{itemize}
There is a problem associated with the electron structure of the superheavy atoms,
in particular, ``relativistic effects" related to the approaching to the speed of light for orbital electrons.
The answer to this question is one of actual problems of atomic physics \cite{Oganessian2017, Eliav2019, Kaygorodovetal2021, Lackenby2018, Mewes, Seif2019}.

The study of superheavy transuranic elements is primarily theoretical in nature due to the extreme complexity of experimental measurements, which are complicated by short half-lives and low production  yields.
In modern treatments based on the Dirac–Coulomb–Breit Hamiltonian, coupled-cluster methods, or density-functional theory, these elements are described with greater accuracy and predictive power (see, for example, \cite{AkhBer, Colloquium2019, FleKar1971, Greineretal1985, Gum2005, Gum2007, Kaygorodovetal2021, Mohretal1998, Oganessian2017, PomSmor1945, ShabQED, ShabQEDHI} and references therein). 
The one- and two-electron systems are an ideal testing ground for addressing complications with electronic shells.
The semiclassical approach, discussed here, may serve as a good starting point with further utilization of more advanced methods
based on the Dirac equation.
This may help to better understand the complexity of the electron shell creation dynamics in superheavy elements.
How will the shell around the nucleus be created? One electron at a time?
In this case, what is the probability that the superheavy nucleus will capture the incoming electron 
and the entire atomic system will disintegrate?
%
As a result, creating a complete electron shell is a complex issue.
{\scshape{Sommerfeld vs Oganesson puzzles\/:}}
As well-known, the energy levels of a hydrogen-like system in Dirac's theory match, precisely, the Sommerfeld fine structure formula (\ref{BidSomEnd}) -- an outcome known as the \textquotedblleft Sommerfeld Puzzle\textquotedblright\ \cite{Biedenharn1983}, discussed further in \cite[pp.~426--429]{Eckert}. This \textquotedblleft puzzle\textquotedblright\ has recently been resolved \cite{SusPuzz}.
Here, in a similar fashion, we rely on the semiclassical model beyond its domain of validity and acknowledge the absence of quantitative quantum electrodynamic (QED) corrections, nonetheless 
providing a new conceptual and pedagogical exploration linking of early quantum mechanics to superheavy ions.
But why does the presence of an extra loop in the `old' Bohr-Sommerfeld, or relativistic classical trajectory of the electron in the transuranic heavy ions under consideration coincide with the parameters of the artificially created element on Earth, Oganesson, which is located at the end of Mendel{\'e}ev's periodic table?
%


\noindent \textbf{Acknowledgments.}
The author would like to express his sincere gratitude to Professor~Ruben~Aba\-gyan, Professor~Krishnaswami Alladi, Dr.~Sergey~I.~Kryuch\-kov, Dr.~Andreas~Ruffing, and Dr.~Eugene~Stepa\-nov for their valuable insights and help. 
Special thanks to Dr.~Kamal~Barley for creation of the original version of the complementary Mathematica notebook.

\appendix
%
\section{Velocity Estimate}
The classical relativistic Hamiltonian, or total energy $E$, of hydrogen-like systems under an
attractive Coulomb potential, has the quadratic form:%
\begin{equation}
\left( E+\frac{Ze^{2}}{r}\right) ^{2}=\mathbf{p}^{2}c^{2}+m^{2}c^{4}, \qquad \mathbf{p} = \gamma m \mathbf{v}.
\label{BidSom1}
\end{equation}%
In the relativistic case, $\gamma =\left(1-v^{2}/c^{2}\right) ^{-1/2}$ is the familiar Lorentz factor \cite[pp.~481--482]{Goldstein} and \cite[pp.~100--102]{LaLif2}. Thus
\begin{equation}
\mathbf{p}^{2}c^{2}+m^{2}c^{4} = \gamma^2 m^2c^4 .
\label{BidSom2}
\end{equation}%
For the ground state, when $n_{r}=0$ and $n_{\theta}=1\/,$ one gets by (\ref{BidSomEnd})--(\ref{BidSom20}) for the circular orbit $\epsilon=0$ and
\begin{equation}
\dfrac{E}{mc^2}=\omega=\dfrac{aZ}{a_0}= \sqrt{1-\alpha^2 Z^2} .
\label{BidSom2}
\end{equation}%
Therefore,
\begin{equation}
 E+\frac{Ze^{2}}{r} =\dfrac{mc^2}{\sqrt{1-\alpha^2 Z^2}} .
\label{BidSom3}
\end{equation}%
As a result, we obtain the estimate (\ref{GroundSpeed}) after simplification.



\begin{thebibliography}{99}
\bibitem{AkhBer} A.~I.~Akhiezer and V.~B.~Berestetskii, \textsl{Quantum
Electrodynamics\/,} Interscience, New York, 1965.

\bibitem{AlladiKlauderRao}  K.~Alladi, J.~R.~Klauder, and C.~R.~Rao, eds., 
\textsl{The Legacy of Alladi Ramakrishnan in the Mathematical Sciences,~\/}
Springer, New York, Dordrecht, Heidelberg, London, 2010.
{\url{https://link.springer.com/book/10.1007/978-1-4419-6263-8}}

\bibitem{AlRamaProb} Alladi Ramakrishnan, {\textsl{Probability and Stochastic Prosesses\/,}}
in: Handbuch der Physik \textbf{3}, pp.~524--651, Springer, Berlin, 1959.
{\url{https://link.springer.com/chapter/10.1007/978-3-642-45912-2_5}}

\bibitem{AlRamaParticles} Alladi Ramakrishnan, {\textsl{Elementary Particles and Cosmic Rays\/,}} 
A Pergamon book, The McMillan Company, New York, 1962.

\bibitem{AlRamaDiracMat} Alladi Ramakrishnan, {\textsl{$L$-matrix Theory Or, The Grammar of Dirac Matrices,\/}}
Tata McGraw-Hill Publishing Company, 1972.

\bibitem{AlRamaDiracSpecialR} Alladi Ramakrishnan, {\textsl{Special Relativity\/,}}
East West Books, Madras, India, 2005.

\bibitem{AlRamaDiary} Alladi Ramakrishnan, {\textsl{Alladi Diary: Memoirs of Alladi Ramakrishnan\/,}} 
K.~Alladi, ed., World Scientific, New Jersey et al., 2019
{\url{https://www.amazon.com/Alladi-Diary-Memoirs-Ramakrishnan-ebook/dp/B07PZK9RQH}}

\bibitem{Barleyetal2021} K.~Barley, J.~Vega-Guzm\'{a}n, A.~Ruffing, and
S.~K.~Suslov, \emph{Discovery of the relativistic Schr\"{o}dinger equation\/}%
, Physics--Uspekhi \textbf{65} no.~1 (2022), 90--103 [in English]. {%
\url{https://iopscience.iop.org/article/10.3367/UFNe.2021.06.039000}}

\bibitem{Barleyetal2025} K.~Barley, A.~Ruffing, and S.~K.~Suslov, \emph{Old
quantum mechanics by Bohr and Sommerfeld from modern perspective\/},
Physics--Uspekhi \textbf{69} no.~1 (2026), 74--93 [in English];
with complementary Mathematica notebooks:
EllipsesAnimateAu.nb, AppendixE.nb, BohrAtomMathematica.nb, and
EllipsesAnimateSofisticated.nb. %
\url{https://ufn.ru/en/articles/2026/1/e/} 

\bibitem{BarleySusMathTwo} K.~K.~Barley, A.~L.~Ruffing, and S.~K.~Suslov,
EllipsesAnimateU92.nb, EllipsesAnimateCn112.nb, EllipsesAnimateOg118.nb,
EllipsesAnimateUbb122.nb, EllipsesAnimateUbe128.nb,
EllipsesAnimateUtb132.nb, EllipsesAnimateUtp135.nb,
EllipsesAnimateSofisticated.nb, complementary Mathematica notebooks.

\bibitem{BarleyRufSusOg} K.~Barley, A.~Ruffing, and S.~K.~Suslov, 
\emph{Oganesson versus Uranium Hydrogen-like Ions and Beyond from the Viewpoint of Old Quantum Mechanics,\/}
 arXiv:2509.06249v4 [quant-ph] 23 Sep 2025. 
 {\url{https://arxiv.org/abs/2509.06249}}

\bibitem{TransU} Complementary Mathematica notebook, Transuranics.nb, about 20~MB.

\bibitem{Biedenharn1983} L.~C.~Biedenharn, \emph{The {\textquotedblleft
Sommerfeld puzzle''} revisited and resolved}, Foundations of Physics \textbf{%
13} no.~1 (1983), 13--34. {%
\url{https://link.springer.com/article/10.1007/BF01889408}}

\bibitem{CeulemansThyssen2018} A.~Ceulemans and P.~Thyssen, \emph{The
``Chemical Mechanics'' of the Periodic Table\/}, in: \textsl{Mendeleev to
Oganesson: A Multidisciplinary Perspective on the Periodic Table\/},
(E.~Scerri and G.~Restrepo, eds.), Oxford University Press, 2018,
pp.~104--121.

\bibitem{Chap2018} K.~Chapman, \emph{The oganesson odyssey\/}, Nature
Chemistry \textbf{10} (2018), 796. {%
\url{https://www.nature.com/articles/s41557-018-0098-4}}

\bibitem{Eckert} M.~Eckert, \textsl{Arnold Sommerfeld. Science, Life and
Turbulent Times,} Springer, New York, Heidelberg, Dordrecht, London, 2013. {%
\url{https://ia800203.us.archive.org/16/items/1461474604_Arnold_Sommerfeld/1461474604_Arnold.pdf}%
}

\bibitem{Eliav2019} E.~Eliav, A.~Borschevsky, and U.~Kaldor, \emph{Electronic structure at the edge of the periodic table,\/}
Nuclear Physics News \textbf{29} (2019)~\#~1, 16--20.
\url{https://doi.org/10.1080/10619127.2019.1571794}

\bibitem{Elyashevich1985} M.~A.~El'yashevich, \emph{Niels Bohr's development
of the quantum theory of the atom and the correspondence principle $($his
1912--1923 work in atomic physics and its significance$)$\/}, Soviet
Physics--Uspekhi \textbf{28} no.~10 (1985), 879--909. 
{\url{https://iopscience.iop.org/article/10.1070/PU1985v028n10ABEH003949}}

\bibitem{Colloquium2019} S.~A.~Giuliani et al., \emph{Colloquium: Superheavy
elements: Oganesson and beyond\/}, Review of Modern Physics \textbf{91}
no.~1 (2019), 011001-1--011001-25 {%
\url{https://journals.aps.org/rmp/abstract/10.1103/RevModPhys.91.011001}}

\bibitem{FleKar1971} G.~N.~Flerov and S.~A.~Karamian, \emph{The search for
superheavy elements in nature: foundations and prospects\/}, in: \textsl{%
Atti Del Accademia Nationale Dei Lincei Convegno Mendeleeviano: Periodicita
E Simmetrie Nella Struttura Elementare Della Materia\/}, (M.~Verde, ed.),
Vincenzo Bona, Turin, 1971, pp.~73--95.

\bibitem{Goldstein} H.~Goldstein, C.~Poole, and J.~Safko, \textsl{Classical
Mechanics,} 3rd edn., Addison Weysley, San Francisco, New York, London,
Toronto, 2000.

\bibitem{Greineretal1985} W.~Greiner, B.~M{\"u}ller, and J.~Rafelski, 
\textsl{Quantum Electrodynamics of Strong Fields: With an Introduction into
Modern Relativistic Quantum Mechanics\/,} Springer-Verlag, Berlin,
Heidelberg, New York, Tokyo, 1985.

\bibitem{Gum2005} A.~Gumberidze et al., \emph{Quantum electrodynamics in
strong electric fields: the ground state Lamb shift in hydrogenlike uranium\/%
}, Physical Review Letters \textbf{94} (2005), 1--4. {%
\url{https://journals.aps.org/prl/abstract/10.1103/PhysRevLett.94.223001}}

\bibitem{Gum2007} A.~Gumberidze et al., \emph{Precision tests of QED in
strong fields: experiments on hydrogen- and helium-like uranium\/}, Journal
of Physics: Conference Series \textbf{58} (2007), 87--92. {%
\url{https://iopscience.iop.org/article/10.1088/1742-6596/58/1/013}}

\bibitem{Hoffman112} S.~Hofmann et al., \emph{The new element 112\/}, Short
note, Zeitschrift f\"{u}r Physik~A \textbf{354} (1996), 229--230. %
\url{https://doi.org/10.1007/BF02769517}

\bibitem{Kaygorodovetal2021} M.~Y.~Kaygorodov et al., \emph{Electron
affinity of oganesson\/}, Physical Review A \textbf{104} (2021), 012819 (10
pages).

\bibitem{Kragh2012} H.~Kragh, \textsl{Niels Bohr and the Quantum Atom: The
Bohr Model of Atomic Structure 1913--1925}, Oxford University Press, Oxford,
2012.

\bibitem{Kragh18} H.~Kragh, \textsl{From Transuranic to Superheavy Elements:
A Story of Dispute and Creation\/}, Springer Briefs in History of Science
and Technology, Springer Nature, Berlin, Cham, Switzerland, 2018.

\bibitem{KraghBohr} H.~Kragh, Ed., \textsl{Niels Bohr on the Constitution of
Atoms and Molecules}, Classic Texts in the Sciences, Birkh\"{a}user,
Springer Nature Switzerland AG, Cham, 2022.

\bibitem{LaLif2} L.~D.~Landau and E.~M.~Lifshitz, \textsl{The Classical
Theory of Fields\/}, 4th edn., Butterworth-Heinemann, Amsterdam--Tokio, 1994.

\bibitem{Lackenby2018} B.~G.~C.~Lackenby, V.~A.~Dzuba, and V.~V.~Flambaum, 
\emph{Atomic structure calculations of superheavy noble element oganesson (Z=118),\/} 
Physical Review A \textbf{98} (2018), 042512 (2018) (5 pages).
    \url{https://journals.aps.org/pra/abstract/10.1103/PhysRevA.98.042512}

\bibitem{Mewes} J.-M.~Mewes, P.~Jerabek  et al., \emph{Oganesson is a semiconductor: 
On the relativistic band-gap narrowing in the heaviest noble-gas solids,\/}
Angewandte Chemie 0.1002/ange.201908327 920190 (2019).
\url{http://dx.doi.org/10.1002/anie.201908327}

\bibitem{Mohretal1998} P.~J.~Mohr, G.~Plunien, G~Soff, \emph{QED corrections
in heavy atoms\/}, Physics Reports \textbf{293} no.~~5-6 (1998), 227--369. {%
\url{https://www.sciencedirect.com/science/article/abs/pii/S037015739700046X}%
}

\bibitem{Myasoedov1985} B.~F.~Myasoedov and N.~Y.~Kremliakova, \emph{Studies of Americium and Curium Solution Chemistry in the USSR,\/} 
in: \textsl{Americium and Curium Chemistry and Technology. Topics in f-Element Chemistry,\/}
(N.~M.~Edelstein, J.~D.~Navratil, and W.~W.~Schulz, Eds), Vol 1. Springer, Dordrecht, pp.~53--79, 1985. 
    \url{https://doi.org/10.1007/978-94-009-5444-1_6}

\bibitem{Oganessian2017} Yu.~Oganessian, \emph{Discovery of the Island of
Stability for Superheavy Elements\/}, Proceedings of IPAC2017, 10 Opening,
Closing and Special Presentations, 03 Special Presentation, Copenhagen,
Denmark, 2017, pp.~4848--4851. {%
\url{https://proceedings.jacow.org/ipac2017/papers/fryaa1.pdf}}

\bibitem{Oganessian2006} Yu.~Ts.~Oganessian et al., \emph{Synthesis of the
isotopes of elements $118$ and $116$ in the $\ ^{249}Cf$ and $\ ^{245}Cm +\
^{48}Ca$ fusion reactions\/}, Physical Review C \textbf{74} (2006), 044602
(9 pages). {%
\url{https://journals.aps.org/prc/abstract/10.1103/PhysRevC.74.044602}}

\bibitem{Oganessian2012} Yu.~Ts.~Oganessian et al., \emph{Production and
decay of the heaviest nuclei $\ ^{293;294}117$ and $\ ^{294}118\/$\/},
Physical Review Letters PRL \textbf{109} (2012), 162501 (5 pages).  {%
\url{https://journals.aps.org/prl/abstract/10.1103/PhysRevLett.109.162501}}

\bibitem{PomSmor1945} I.~Pomeranchuk and J.~Smorodinsky, \emph{On the energy
levels of systems with $Z>137$\/}, Journal of Physics USSR \textbf{9} no.~2
(1945), 97--100.

\bibitem{Powers1971} R.~T.~Powers, \emph{Frequencies of Radial Oscillation
and Revolution as Affected by Features of a Central Potential\/}, in: 
\textsl{Atti Del Accademia Nationale Dei Lincei Convegno Mendeleeviano:
Periodicita E Simmetrie Nella Struttura Elementare Della Materia\/},
(M.~Verde, ed.), Vincenzo Bona, Turin, 1971, pp.~235--242.

\bibitem{Reed} B.~C.~Reed, {\textsl{The Bohr Atom: A Guide,}} IOP
Publishing, Bristol, UK, 2020.

\bibitem{SchrQMI} E.~{Schr{\"{o}}dinger}, \emph{Quantisation as a problem of
proper values $($Part~I$)$,} in: \textsl{Collected Papers on Wave Mechanics}%
, New York, Providence, Rhode Island: AMS Chelsea Publishing, 2010,
pp.~1--12; [German Original: Annalen der Physik (4), vol. 79(6),
pp.~489--527, 1926)].

\bibitem{ShabQED} V.~M.~Shabaev, \emph{Two-time Green's function method 
in quantum electrodynamics of high-$Z$ few-electron atoms\/},
Physics Reports \textbf{356} (2002)~\#~3, 119--228.
{\url{https://www.sciencedirect.com/science/article/abs/pii/S0370157301000242?via%3Dihub}}

\bibitem{ShabQEDHI} V.~M.~Shabaev, \emph{Quantum electrodynamics of heavy ions and atoms: 
current status and prospects\/}, Physics--Uspekhi \textbf{51} no.~11 (2008), 1175--1180.
{\url{https://iopscience.iop.org/article/10.1070/PU2008v051n11ABEH006801/meta}}

\bibitem{Seif2019} W.~M.~Seif, H.~Anwer, A.~R.~Abdulghany, \emph{Ground-state and stability properties of $\ {^{288-308}_{118}}$Og
isotopes based on semi-microscopic calculations,\/}
Annals of Physics \textbf{401} (2019) 149--161.
\url{https://www.sciencedirect.com/science/article/abs/pii/S0003491618303208?via%3Dihub}

\bibitem{Somm1916} A.~Sommerfeld, \emph{Zur Quantentheorie der Spektrallinien%
}, Annalen der Physik \textbf{51} no.~17 (1916), 1--94 [in German]. {%
\url{https://onlinelibrary.wiley.com/doi/10.1002/andp.19163561702}}

\bibitem{SomAS} A.~Sommerfeld, \textsl{Atomic Structure and Spectral Lines},
Volume I, 3rd English edn., E.~P.~Dutton and Company Inc. Publishers., New
York, 1936. {%
\url{https://archive.org/details/atomicstructures0000somm/page/n7/mode/2up}}%
\,; 2nd German edn. {%
\url{https://archive.org/details/atombauundspekt00sommgoog/page/478/mode/2up}%
}.

\bibitem{SusRelInt} S.~K.~Suslov, \emph{Expectation values in relativistic
Coulomb problems\/}, Journal of Physics B: Atomic, Molecular and Optical
Physics \textbf{42} no.~18 (2009), 185003. {%
\url{https://iopscience.iop.org/article/10.1088/0953-4075/42/18/185003}}

\bibitem{SusPuzz} S.~K.~Suslov, \emph{The \textquotedblleft Sommerfeld's
puzzle\textquotedblright\ and its extensions}, in: \textsl{The Proceedings
of Second International Workshop on Quantum Nonstationary Systems},
(A.~Dodonov and C.~C.~H.~Ribeiro, Eds.), LF Editorial, S\~{a}o Paulo, pp.~43--59,  2024. {%
\url{https://lfeditorial.com.br/produto/proceedings-of-the-second-international-workshop-on-quantum-nonstationary-systems/}%
}


\bibitem{Wheeler1971} J.~A.~Wheeler, \emph{From Mendel{\'e}ev's Atom to the Collapsing Star\/}, 
in: \textsl{Atti Del Accademia Nationale Dei Lincei Convegno Mendeleeviano:
Periodicita E Simmetrie Nella Struttura Elementare Della Materia\/}, (M.~Verde, ed.), 
Vincenzo Bona, Turin, 1971, pp.~189--223;
see also {\em Transactions of the New York Academy of Sciencesem, 1971,\/}
  {\bf 33}(8), 745--779
  \url{https://doi.org/10.1007/978-94-010-2126-5_15}
  \url{https://nyaspubs.onlinelibrary.wiley.com/doi/epdf/10.1111/j.2164-0947.1971.tb02638.x}

\end{thebibliography}
\end{document}